# Antenne à métasurface scalaire de polarisation linéaire

A.Arroyo[1], R.Contreres[2], A.Piche[3], H.Roussel[1], M.Casaletti[1]

[1] Sorbonne Université, CNRS, Laboratoire de Génie Electrique et Electronique de Paris, 75252 - Paris, France.
alejandro.arroyo, helene.roussel, massimiliano.casaletti@sorbonne-universite.fr
Université Paris-Saclay, CentraleSupélec, CNRS, Laboratoire de Génie Electrique et Electronique de Paris,
91192 - Gif-sur-Yvette, France.
[2] CNES : Radiofrequency department, Antenna Team, Toulouse, France. romain.contreres@cnes.fr
[3] Airbus Defence and Space S.A.S, Toulouse, France. alexandre.piche@airbus.com

***Résumé*** – Une nouvelle approche utilisant des métasurfaces scalaires pour la conception d'antennes à polarisation linéaire est présentée. La méthode proposée est basée sur la construction de l'impédance de surface $Z_s$ avec une technique dite de « phase-matching », utilisant la somme de deux polarisations circulaires en opposition de phase. Ce processus permet d'obtenir de bonnes performances de l'antenne synthétisée, telles que la réduction du niveau des lobes secondaires et l'obtention d'un lobe principal presque symétrique quelle que soit la direction de pointage. Des résultats numériques et de mesures sont également présentés.

## 1. Introduction

Les antennes à métasurface se présentent de nos jours comme de bonnes solutions pour nos systèmes de communications. Ces antennes ont démontré à plusieurs reprises qu'elles pouvaient être associées à de bonnes performances pour différentes applications [1], [2], [3]. Le principe général de ces antennes utilise le phénomène de conversion d'onde de surface en onde de fuite, en modulant la métasurface de façon adéquate.

Ainsi, une polarisation linéaire a été obtenue en [4] et en [5] en subdivisant la métasurface en deux régions équivalentes et en opposition de phase.

Ici une nouvelle approche pour l'obtention de polarisions linéaires à base de métasurfaces scalaires est présentée. Elle utilise une somme de deux polarisations circulaires inverses réalisée par la construction d'une double modulation spirale. Cette nouvelle approche permet une meilleure transition dans le gradient des éléments de base. Ces travaux peuvent être extrapolés à d'autres géométries de base scalaire pour la métasurface. La géométrie des éléments de base utilisés ici sont des motifs carrés métalliques imprimés de façon non uniforme sur un substrat diélectrique. Nous présenterons également des résultats de simulations.

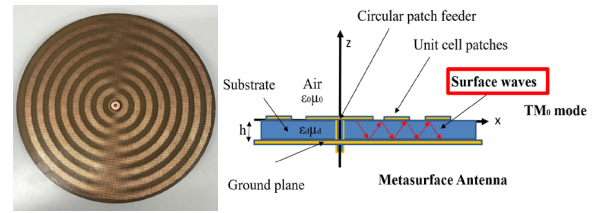

**Fig. 1.** Géométrie de l'antenne: antenne imprimée et schéma de l'antenne vue de profil. La source coaxiale permet de générer une onde de surface $TM_0$.

## 2. Théorie

La géométrie générale de l'antenne à métasurface est montrée sur la Fig. 1. La métasurface est composée d'éléments de base carrés métalliques placés sur un substrat diélectrique de permittivité relative $\varepsilon_r$ et de hauteur $h$). Cette hauteur $h$ ainsi que l'impédance moyenne $\bar{X}_s$ ont été choisies afin d'assurer la seule propagation du mode dominant $TM_0$ [6]. A l'interface ($z=0$), les conditions aux limites établissent la relation entre les champs tangentiels et le tenseur d'impédance de surface :

$$\mathbf{E}_t = Z_s \hat{\mathbf{n}} \times \mathbf{H}_t = Z_S \mathbf{J} \quad (1)$$

Où $\hat{\mathbf{n}}$ est le vecteur normal à la surface, l'indice t indique tangentiel et **J** est le courant surfacique électrique. L'onde de surface $TM_0$ est excitée par l'alimentation centrale et sa constante de propagation $k_{sw}^t$ est donnée par :

$$k_{sw}^t = k_0\sqrt{1+\left(\bar{X}_s(\hat{\mathbf{k}}_{sw}^t)/\zeta\right)^2} \quad (2)$$

Où $k_0$ est la constante de propagation de l'espace libre, $\zeta$ l'impédance de l'espace libre et $\bar{X}_s$ la réactance surfacique moyenne de la métasurface.

Une onde de fuite se produit si la quantité $k_{sw}^t(\rho')-2\pi/p_{ij}(\rho')$ est inférieure à $k_0$ (le mode -1 de l'expansion de Floquet-Bloch) [5] ; le courant surfacique magnétique équivalent $\mathbf{J}_M^{(-1)}(\rho')$ est alors donné par :

$$\mathbf{J}_M^{(-1)}(\rho') = \bar{X}_s M(\rho') e^{-j\left(k_{sw}^t - \frac{2\pi}{p(\rho')}\right)\hat{\mathbf{k}}_{sw}^t(\rho')\cdot\rho'} \hat{\boldsymbol{\phi}} \quad (3)$$

Où une modulation de type sinusoïdale de l'impédance surfacique $Z_s$ a été appliquée. Les paramètres de modulation sont : la réactance surfacique moyenne notée $\overline{X}_s$, l'indice de modulation $M$ et la périodicité $p$).

Si les conditions physiques énoncées précédemment sont vérifiées alors une onde de fuite peut alors être rayonée à travers cette surface.

## 3. Polarisations linéaires

La source d'alimentation placée au centre de l'antenne excite une onde de surface cylindrique donnée par :

$$\mathbf{H}_t(\rho') = A(\rho')\mathrm{H}_1^{(2)}(k_t\rho)\hat{\phi} \quad (4)$$

Le choix de la polarisation d'émission se déduit en appliquant la théorie de champs d'ouverture lié à la distribution du courant magnétique. Une polarisation circulaire dans une direction $(\theta_0, \phi_0)$ est obtenue utilisant la distribution de courant surfacique [6] :

$$\mathbf{J}_\mathrm{M}^{CP}(\rho') = e^{\pm j\phi'}e^{-jk_0\rho'\sin\theta_0\cos(\phi'-\phi_0)}\hat{\phi} \quad (5)$$

Où le signe ± indique le sens de la polarisation circulaire (CP) droite ou gauche.

On obtient une polarisation linéaire (LP) grâce à la somme cohérente de deux polarisations circulaires pour obtenir :

$$\mathbf{J}_\mathrm{M}^{LP}(\rho') = \begin{pmatrix} e^{+j\phi'}e^{-jk_0\rho'\sin\theta_0\cos(\phi'-\phi_0)} \\ \pm e^{-j\phi'}e^{-jk_0\rho'\sin\theta_0\cos(\phi'-\phi_0)} \end{pmatrix}\hat{\phi} \quad (6)$$

Pour aboutir à la construction de l'impédance de surface $Z_s$, on impose une équivalence entre (3) et (6) :

$$\mathbf{J}_\mathrm{M}^{(-1)}(\rho') \propto \mathbf{J}_\mathrm{M}^{LP}(\rho') \quad (7)$$

Finalement la métasurface scalaire est conçue en sélectionnant localement l'impédance $Z_s$ en termes d'amplitude (réactance moyenne $\overline{X}_s$ et indice de modulation $M$) et de phase pour la polarisation linéaire souhaitée (périodicité $p$).

## 4. Validation

### 4.a. Simulation numérique

Une antenne à métasurface polarisée linéairement selon la méthode décrite ci-dessus a été conçue et simulée. A la fréquence $f = 20\ GHz$, cette antenne rayonne un faisceau dépointant dans la direction $\theta=35°$ et $\phi=0°$, avec une polarisation linéaire attendue selon $\vec{u}_\theta$. L'onde de surface incidente est générée par un patch circulaire alimenté par un câble coaxial placé au centre.

Les éléments de base de la métasurface consistent en carrés métalliques imprimés sur un substrat FR4 de permittivité relative $\varepsilon_r = 4.3$, angle de pertes $\tan\delta = 0.02$ et de hauteur $h=1.6mm$. Leurs dimensions sont choisies en appliquant la procédure décrite dans la section 3. L'antenne finale est de rayon 6 $\lambda_0$ (longueur d'onde dans l'espace libre).

La Fig. 2 montre le champ électrique pour les composantes en co-polarisation (courbe bleu) et cross-polarisation (courbe rouge), simulés avec le logiciel commercial CST dans le plan $\phi=0°$.

Comme attendu, un faisceau dépointant à $\theta=35°$ est obtenu, avec une polarisation croisée assez faible et un lobe secondaire à -10dB. Le niveau élevé de ce lobe est dû au dépointage.

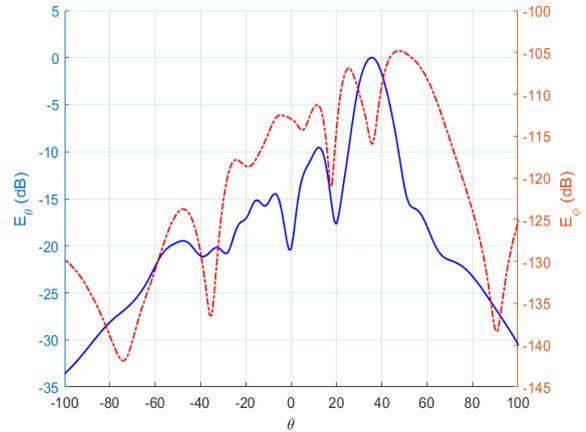

**Fig. 2.** Champ Electrique rayonné simulé en co-polarisation et cross-polarisation à 20GHz dans le plan $\phi=0°$.

### 4.b. Simulation numérique et validation expérimentale

Une autre antenne à métasurface polarisée linéairement selon la méthode décrite ci-dessus a été conçue et mesurée expérimentalement en chambre anéchoïque. Cette antenne rayonne, à la fréquence $f = 20\ GHz$, un faisceau dépointant dans la direction $\theta=0°$ et $\phi=0°$, avec une polarisation linéaire attendue selon $\vec{u}_\theta$. L'onde de surface incidente est générée par un patch circulaire alimenté par un câble coaxial placé au centre, comme précédemment. La Fig. 3 montre le champ électrique en fonction de l'angle $\theta$ pour les composantes en co-polarisation (courbe bleu) et cross-polarisation (courbe rouge), simulé avec le logiciel commercial CST dans le plan $\phi=0°$, ainsi que la mesure réalisée dans la base de mesure du GeePs (pointillé noir). Comme attendu, un faisceau dépointant à $\theta=0°$ est obtenu, avec une polarisation croisée assez faible et un lobe secondaire à -10dB.

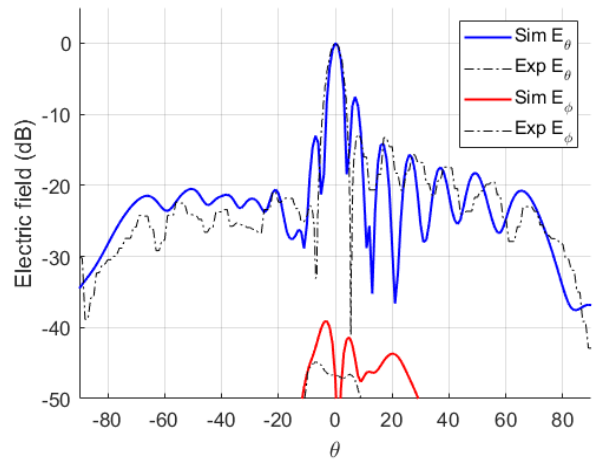

**Fig. 3.** Champ Electrique rayonné simulé et mesuré en co-polarisation et cross-polarisation à 20GHz dans le plan $\phi=0°$.

## 5. Conclusion

Une antenne à métasurface polarisée linéairement a été présentée. Pour cela nous utilisons une nouvelle méthode de conception résultant de deux polarisations circulaires inverses obtenues en réalisant une double modulation spirale. Cette approche permet une meilleure transition dans la sélection des éléments de base, aboutissant ainsi à la formation d'un faisceau plus symétrique que l'antenne proposée dans [5]. Les résultats de simulations sont validés par des mesures réalisées en chambre anéchoïque. Cette même procédure peut être étendue et appliquée à n'importe quelle géométrie de métasurface scalaire.

## Références